\documentclass[11pt]{article}

\usepackage{amsmath}
\usepackage{graphicx}
\usepackage{amsfonts}
\usepackage{amssymb}
\usepackage{epsfig}
\usepackage{color}
\usepackage{psfrag}
\usepackage{epstopdf}

\newcommand{\EQ}{\begin{equation}}
\newcommand{\EN}{\end{equation}}
\newcommand{\be}{\begin{equation}}
\newcommand{\ee}{\end{equation}}
\newcommand{\bea}{\begin{eqnarray}}
\newcommand{\eea}{\end{eqnarray}}

\begin{document} \setcounter{page}{0}
\topmargin 0pt
\oddsidemargin 5mm
\renewcommand{\thefootnote}{\arabic{footnote}}
\newpage
\setcounter{page}{0}
\topmargin 0pt
\oddsidemargin 5mm
\renewcommand{\thefootnote}{\arabic{footnote}}
\newpage
\begin{titlepage}
\begin{flushright}
SISSA 40/2012/EP \\
\end{flushright}
\vspace{0.5cm}
\begin{center}
{\large {\bf Parafermionic excitations and critical exponents}\\
{\bf  of random cluster and $O(n)$ models}}\\
\vspace{1.8cm}
{\large Gesualdo Delfino}\\
\vspace{0.5cm}
{\em SISSA -- Via Bonomea 265, 34136 Trieste, Italy}\\
{\em INFN sezione di Trieste}\\
\end{center}
\vspace{1.2cm}

\renewcommand{\thefootnote}{\arabic{footnote}}
\setcounter{footnote}{0}

\begin{abstract}
\noindent
We introduce the notion of parafermionic fields as the chiral fields which describe particle excitations in two-dimensional conformal field theory, and argue that the parafermionic conformal dimensions can be determined using scale invariant scattering theory. Together with operator product arguments this may provide new information, in particular for non-rational conformal theories. We obtain in this way the field theoretical derivation of the critical exponents of the random cluster and $O(n)$ models, which in the limit of vanishing central charge yield percolation and self-avoiding walks. A simple derivation of the relation between $S$-matrix and Lagrangian couplings of sine-Gordon model is also given.
\end{abstract}
\end{titlepage}

\newpage

\section{Introduction}
Field theory is the natural framework for the study of universal properties of statistical models exhibiting continuous phase transitions. The two-dimensional case remains exceptional to date because it allows for exact solutions. In particular, two-dimensional conformal field theory \cite{BPZ} yields a classification of renormalization group fixed points which can be associated to critical points of statistical models through symmetry arguments, with the determination of critical exponents as a main consequence. 

In two dimensions exact solvability is often possible already on the lattice \cite{Baxter} and many critical exponents provided by conformal field theory were first obtained within some discrete formulation of the given universality class. There are, however, examples for which the lattice derivation has been the only one available to date. This is the case of the two basic universality classes associated to $S_q$ (permutational) and $O(n)$ symmetry. The first one is represented on the lattice by the $q$-state Potts model, which allows for a representation with $q$ real corresponding to a model of random clusters \cite{FK}, with ordinary percolation arising as $q\to 1$. The $O(n)$ model  also allows for continuation to real values of $n$ and gives self-avoiding walks in the limit $n\to 0$ \cite{deGennes}.

The critical exponents of the Potts and $O(n)$ models have been obtained within the framework of the lattice Coulomb gas (see \cite{Nienhuis} for a review). This consists of a number of mappings between lattice models leading to an identification with a system of interacting electric and magnetic charges, which in turn can be related by renormalization group arguments to a Gaussian model for which exponents are computable. While this non-trivial procedure makes a field theoretical derivation highly desirable, the continuation to real values of $q$ or $n$ leads to non-rational conformal field theories for which a canonical conformal description is not available. This is why, when the lattice results for the critical exponents were cast into the conformal language in \cite{Dotsenko,DF}, the essential issue of the dependence on the parameters $q$ and $n$ remained as an external input inferred from the lattice results. 

In this paper we give a purely field theoretical derivation of the critical exponents of the $S_q$ and $O(n)$ models, independent of any lattice input. This is obtained as a consequence of a more general observation about the relation between the spin of the fields associated to the critical particle excitations (which in general is fractional, justifying the name of parafermionic fields) and scale invariant scattering amplitudes. Together with the fact that we are able to determine the parafermionic spin from operator product expansion and mutual locality arguments, and the scattering amplitudes from the enforcement of $S_q$ or $O(n)$ symmetry, the above mentioned relation leads to a relatively simple determination of the exponents.

The paper is organized as follows. In the next section we recall the notion of mutual locality and introduce parafermions and their scattering properties. Section~3 is devoted to conformal field theory with central charge $c\leq 1$, in particular to characterize the critical lines with $c<1$ with the properties required for the purposes of this paper. The critical scattering theories with $S_q$ or $O(n)$ symmetry are then solved in section~4. Section~5 summarizes the results and contains some additional comments.

\section{Fields and particles at criticality}
Consider a spin model of classical statistical mechanics which undergoes a continuous phase transition when a temperature parameter $T$ takes a critical value $T_c$. Within the field theory describing the universal properties of the scaling limit we call energy field $\varepsilon(x)$ the field conjugated to the parameter $\tau\sim T-T_c$; we denote by $x=(x_1,x_2)$ a point in two-dimensional Euclidean space. Typically the phase transition arises from the spontaneous breaking of an internal symmetry and the energy field is invariant under the action of the symmetry. The spin (or order parameter) field $\sigma(x)$, on the other hand, transforms according to a representation of the symmetry and has in general several components (we drop indices for the time being). 

The fields $\varepsilon$ and $\sigma$ are examples of scaling fields. In two dimensions a scaling field $\Phi$ is characterized by the conformal dimensions $\Delta_\Phi$ and $\bar{\Delta}_\Phi$, which determine the scaling dimension $X_\Phi=\Delta_\Phi+\bar{\Delta}_\Phi$ and the spin $s_\Phi=\Delta_\Phi-\bar{\Delta}_\Phi$. The field $\Phi$ picks up a phase $e^{-is_\Phi\theta}$ under a rotation by an angle $\theta$ on the Euclidean plane. The scaling dimensions of $\varepsilon$ and $\sigma$ (which are scalar, i.e. spinless, fields) determine the correlation length and magnetization critical exponents as $\nu=1/2(1-\Delta_\varepsilon)$ and $\beta=2\Delta_\sigma\nu$, respectively; the other canonical exponents in zero external field are obtained from the usual scaling relations.

The operator product expansion (OPE) of two scaling fields can be written as
\EQ
\Phi_i(x)\Phi_j(0)=\sum_k C_{ij}^{k}\,z^{\Delta_{k}-\Delta_{i}-\Delta_{j}}\bar{z}^{\bar{\Delta}_{k}-\bar{\Delta}_{i}-\bar{\Delta}_{j}}\,\Phi_k(0)\,,
\label{ope}
\EN
where $\Delta_i\equiv\Delta_{\Phi_i}$, $\{\Phi_k\}$ is a basis of scaling fields, $C_{ij}^{k}$ are structure constants, and $z$ and $\bar{z}$ are related to the Euclidean coordinates as $z=x_1+ix_2$ and $\bar{z}=x_1-ix_2$. The fields $\Phi_i$ and $\Phi_j$ are said to be mutually local if the correlation functions containing the product $\Phi_i(x)\Phi_j(0)$ do not possess a branch point at $x=0$, i.e. are single valued under the continuation $z\to e^{2i\pi}z$, $\bar{z}\to e^{-2i\pi}\bar{z}$. This is the case if the condition
\EQ
\gamma_{ij}^k\equiv s_k-s_i-s_j\in{\bf Z}
\label{gamma}
\EN
holds for all $\Phi_k$ giving a non-zero contribution to the r.h.s. of (\ref{ope}). The fields having a direct physical meaning for the statistical model should be local with respect to the energy field $\varepsilon(x)$.

\begin{figure}
\begin{center}
\includegraphics[width=7cm]{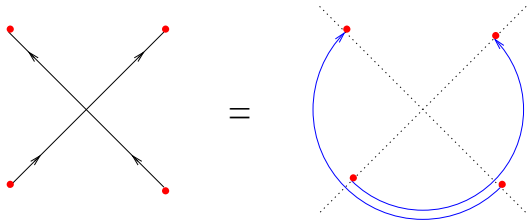}
\caption{Graphical illustration of equation (\ref{phase}).}
\label{paraf_phases}
\end{center} 
\end{figure}

It is often useful to look at one of the Euclidean dimensions as imaginary time. Rotational invariance of the two-dimensional Euclidean system then becomes relativistic invariance from the (1+1)-dimensional point of view. At criticality, the elementary excitations of a relativistic (1+1)-dimensional theory are massless particles with sign of the momentum that cannot be changed by a Lorentz transformation, i.e. they are either right movers or left movers. These particles are described by chiral (or holomorphic) fields $\psi$, with $\bar{\Delta}_\psi=0$, if right movers, and $\bar{\psi}$, with $\Delta_{\bar{\psi}}=0$, if left movers; one also has $\Delta_{{\psi}}=\bar{\Delta}_{\bar{\psi}}=s_\psi=-s_{\bar{\psi}}$. In a scale invariant theory right and left degrees of freedom are decoupled, so that if we consider the scattering of a right mover with a left mover there is no dynamical phase shift\footnote{Scale invariance prevents a momentum dependence of the scattering amplitude. The only relativistic invariant is the center of mass energy and there is no way of building a dimensionless scaling variable.}. Right-left scattering, however, involves exchanging the position on the line of two identical particles, so that scattering entails in general a statistical factor. Since in absence of dynamical interaction the passage from the initial to the final state can also be realized by $\pi$-rotations (see Fig.~\ref{paraf_phases}), the right-left scattering (statistical) phase can be written as
\EQ
S=e^{-i\pi(s_\psi-s_{\bar{\psi}})}=e^{-2i\pi\Delta_\psi}\,.
\label{phase}
\EN
Notice that $\Delta_\psi$ integer (half integer) yields bosonic (fermionic) statistics; the spin, however, can take more general values and for this reason we refer to $\psi$ and $\bar{\psi}$ as parafermionic fields. 

In general parafermionic excitations and the corresponding fields form multiplets transforming under some representation of the internal symmetry of the theory. We denote by $\Psi_a$ the components of such multiplets, and by $\bar{\Psi}_a$ their anti-chiral counterparts. We can exploit the fact that in right-left scattering the two particles exchange their relative positions on the line to encode the most general process in the symbolic relation
\EQ
\Psi_a\circ\bar{\Psi}_b=S_{ab}^{cd}\,\,\,\bar\Psi_d\circ\Psi_c\,,
\label{algebra}
\EN
where $S_{ab}^{cd}$ are scattering amplitudes and summation over repeated indices on the r.h.s. accounts for the fact that in general the parafermionic components may not be scaling fields. In such a case, however, it should be possible to find a linear combination which diagonalizes (\ref{algebra}) yielding a scattering phase and, using (\ref{phase}), a definite conformal dimension.

The scattering amplitudes are subject to the constraints of unitarity and crossing symmetry, which in the present case simply read
\EQ
S_{ab}^{ef}\left[S_{ef}^{cd}\right]^*=\delta_a^c\delta_b^d\,,
\label{unitarity}
\EN
\EQ
S_{ab}^{cd}=\left[S_{a\bar{d}}^{c\bar{b}}\right]^*\,;
\label{crossing}
\EN
the asterisk denotes complex conjugation and the bar over indices denotes antiparticles.

We will illustrate in the next sections that $\Delta_\psi$ can also be determined from the requirement of mutual locality between $\psi$ and $\varepsilon$, and that comparison with (\ref{phase}) may provide non-trivial information.

\section{Conformal field theory with $c\leq 1$}
We will be interested for our purposes in investigating some general properties of conformal field theories with central charge $c\leq 1$. A way of doing this is to consider the conformal field theory of a free scalar boson with action
\EQ
{\cal A}=\frac{1}{4\pi}\int d^2x\,(\partial_a\varphi)^2
\label{action}
\EN
and components of the stress tensor given by
\EQ
T(z)=-(\partial\varphi)^2+iQ\,\partial^2\varphi\,,\hspace{1cm}
\bar{T}(\bar{z})=-(\bar{\partial}\varphi)^2+iQ\,\bar{\partial}^2\varphi\,,
\label{T}
\EN
with $\partial$ and $\bar{\partial}$ denoting derivatives with respect to $z$ and $\bar{z}$, respectively. The terms proportional to $Q$ introduce a modification of the stress tensor which does not change energy and momentum. They affect, however, the central charge of the theory and the conformal dimensions of a field $\Phi$, which appear as coefficients in the OPE of the stress tensor with itself and with $\Phi$, respectively \cite{BPZ}. The central charge becomes \cite{DF}
\EQ
c=1-6Q^2\,.
\label{c}
\EN
Introducing the decomposition $\varphi(x)=\phi(z)+\bar{\phi}(\bar{z})$ allowed by the equation of motion $\partial\bar{\partial}\varphi=0$, the primary fields correspond to the exponentials 
\EQ
V_{p,\bar{p}}=e^{2i(p\phi+\bar{p}\bar{\phi})}\,,
\label{V}
\EN
 with conformal dimensions $(\Delta_p,\Delta_{\bar{p}})$, and \cite{DF}
\EQ
\Delta_p=p(p-Q)\,.
\label{deltap}
\EN

\noindent
{\bf c=1.} It will be useful to start with the case $Q=0$, namely the usual Gaussian fixed point theory with central charge $c=1$ and OPE
\EQ
V_{p_1,\bar{p}_1}\cdot V_{p_2,\bar{p}_2}=\left[V_{p_1+p_2,\bar{p}_1+\bar{p}_2}\right]\,,
\label{gaussian}
\EN
where we omit for simplicity coordinate dependence and structure constants; the square brackets indicate that, besides the primary field, the r.h.s. contains its descendants, with conformal dimensions exceeding those of the primary by positive integers.

A generic choice for the energy field in this theory is $\varepsilon=V_{b,b}+V_{-b,-b}\propto\cos 2b\varphi$ with $\Delta_\varepsilon=b^2$. Using (\ref{gaussian}), (\ref{gamma}) and (\ref{deltap}), we obtain that the fields $V_{p,\bar{p}}$ are local with respect to $\varepsilon$ if $\Delta_{p\pm b}-\Delta_{\bar{p}\pm b}-\Delta_p+\Delta_{\bar{p}}=\pm 2b(p-\bar{p})$ is an integer, i.e. if 
\EQ
p-\bar{p}=\frac{m}{2b}\,,\hspace{1cm}m\in{\bf Z}\,.
\label{quantization}
\EN
It becomes transparent in the fermionized (Thirring) version of the theory \cite{Coleman,Mandelstam} that the integer $m$ is the $U(1)$ fermionic charge. The fundamental parafermions will be the chiral fields local with respect to $\varepsilon$ with the lowest charge, i.e. $\Psi_\pm=V_{\pm 1/2b,0}$, $\bar{\Psi}_\pm=V_{0,\pm 1/2b}$, with $\Delta_{\Psi_\pm}=1/4b^2$. On the other hand, the spin field components associated to the $U(1)\sim O(2)$ symmetry are selected picking up the scalar fields with $m=\pm 1$, i.e. $\sigma_\pm=V_{\pm 1/4b,\mp 1/4b}$ with $\Delta_{\sigma_\pm}=1/16b^2$.

We see in particular that at $b^2=1/2$ the parafermions become ordinary spin $1/2$ fermions, so that this is the point where $c=1$ results from two decoupled Ising models, each with $c=1/2$, energy $\varepsilon_i$ and spin $\sigma_i$ ($i=1,2$). We have $\Delta_\varepsilon=\Delta_{\varepsilon_i}=1/2$ and $\Delta_{\sigma_\pm}=\Delta_{\sigma_1\sigma_2}=2\Delta_{\sigma_i}=1/8$.

\vspace{.3cm}
\noindent
{\bf c$<$1.} When $Q$ is real the central charge (\ref{c}) is smaller than 1. It is useful to define the parameters $\beta$ and $t$ through the relations
\EQ
Q=\beta^{-1}-\beta\,,\hspace{1cm}\beta^2=t/(t+1)\,,
\label{beta-t}
\EN
so that, in particular, $c=1-\frac{6}{t(t+1)}$. We also introduce the notations
\EQ
p_{\mu,\nu}=\frac12[(1-\mu)\beta^{-1}-(1-\nu)\beta]\,,
\label{pmunu}
\EN
\EQ
\Phi_{\mu,\nu}(z)=V_{p_{\mu,\nu},0}\,,\hspace{1cm}\bar{\Phi}_{\mu,\nu}(\bar{z})=V_{0,p_{\mu,\nu}}\,;
\label{phi}
\EN
it follows from (\ref{deltap}) that the single non-vanishing conformal dimension of the fields (\ref{phi}) is
\EQ
\Delta_{\mu,\nu}=\frac{[(t+1)\mu-t\nu]^2-1}{4t(t+1)}\,.
\label{deltamunu}
\EN
It is a fundamental result of $c<1$ conformal field theory that when the indices $\mu$ and $\nu$ take positive integer values $m$ and $n$ the fields (\ref{phi}) become {\it degenerate}\footnote{Rational values of $\beta^2$ yield rational conformal theories that can be consistently built on a finite number of degenerate primaries and their descendants \cite{BPZ}.}, and are characterized, in particular, by the degenerate-nondegenerate and degenerate-degenerate OPE's \cite{BPZ}
\bea
\Phi_{m,n}\cdot\Phi_{\mu,\nu}&=&\sum_{k=0}^{m-1}\,\sum_{l=0}^{n-1}\left[\Phi_{\mu-m+1+2k,\nu-n+1+2l}\right]\,,
\label{dnd}\\
\Phi_{m_1,n_1}\cdot\Phi_{m_2,n_2}&=&\sum_{k=0}^{min(m_1,m_2)-1}\,\sum_{l=0}^{min(n_1,n_2)-1}\left[\Phi_{|m_1-m_2|+1+2k,|n_1-n_2|+1+2l}\right]\,;
\label{dd}
\eea
similar relations hold for the fields $\bar{\Phi}_{\mu,\nu}$.

The energy field $\varepsilon$ is consistently defined if the OPE $\varepsilon\cdot\varepsilon$ does not produce fields with scaling dimension smaller than $X_\varepsilon$. It does not seem possible to satisfy this condition if $\varepsilon$ is non-degenerate, and it can be checked that the only degenerate fields which satisfy this condition, are relevant (in the renormalization group sense, namely have $\Delta<1$) and have positive dimension in the interval $0<c<1$ are $\Phi_{1,2}$, $\Phi_{2,1}$ and $\Phi_{1,3}$. We will confine our discussion to theories which for $c=1/2$ ($t=3$) reduce to the Ising model with $\Delta_\varepsilon=1/2$; then we focus on $\Phi_{2,1}$ and $\Phi_{1,3}$ which indeed satisfy this requirement.

Let us start from $\varepsilon=\Phi_{2,1}\bar{\Phi}_{2,1}$ and look for the corresponding parafermion $\psi(z)$. One can use (\ref{dnd}) and (\ref{gamma}) to check that there are no non-degenerate chiral fields local with respect to $\varepsilon$. On the other hand, for a degenerate chiral field (\ref{dd}) gives 
\EQ
\varepsilon\cdot\Phi_{m,n}=\sum_{k=0}^{min(2,m)-1}\left[\Phi_{|m-2|+1+2k,n}\bar{\Phi}_{2,1}\right]\,,
\EN
and (\ref{gamma}) becomes
\bea
\Delta_{|m-2|+1+2k,n}-\Delta_{2,1}-\Delta_{m,n}&=&
-\frac{1}{2\beta^2}[2m-2k(1+k)-|m-2|(1+2k)-1]\nonumber\\
&&+\frac{1}{2}[1-n(1+2k-m+|m-2|)]\in{\bf Z}\,.
\eea
These conditions are satisfied for any $\beta$ if $m=1$ and $n$ is odd; the requirement $\Delta_\psi=1/2$ at $t=3$ finally fixes $n=3$, i.e. $\psi=\Phi_{1,3}$.

When looking for the spin field it is natural to require for generic $c$ the condition 
\EQ
\varepsilon\cdot\sigma=\sigma+\cdots
\label{epsilonsigma}
\EN
which holds in the Ising case. Since (\ref{dnd}) gives
$\Phi_{2,1}\cdot\Phi_{\mu,\nu}=\left[\Phi_{\mu-1,\nu}\right]+\left[\Phi_{\mu+1,\nu}\right]$, (\ref{epsilonsigma}) is satisfied if $\sigma=\Phi_{1/2,0}\bar{\Phi}_{1/2,0}$ and we identify the fields $\Phi_{\mu,\nu}$ and $\Phi_{-\mu,-\nu}$, which have the same dimension. The OPE (\ref{epsilonsigma}) also generates the scalar field $\Phi_{3/2,0}\bar{\Phi}_{3/2,0}$ which is naturally identified as a subleading spin field; more generally, iteration of the OPE with $\varepsilon$ produces the family of subleading spin fields $\sigma_k=\Phi_{k+1/2,0}\bar{\Phi}_{k+1/2,0}$, $k=1,2,\ldots$. On the other hand, $\varepsilon\cdot\varepsilon$ and iterations produce\footnote{The requirement that $\varepsilon$ is local with respect to itself implies that only scalar primaries and their descendants are actually produced in the OPE $\varepsilon\cdot\varepsilon$. A similar prescription has to be adopted for the OPE's of the fields $\sigma_k$ and $\varepsilon_k$ with $\varepsilon$.} the subleading thermal fields $\varepsilon_k=\Phi_{k+2,1}\bar{\Phi}_{k+2,1}$, $k=1,2,\ldots$. The structure of the OPE is invariant under the transformation $\varepsilon_k\to(-1)^{k+1}\varepsilon_k$, including $\varepsilon_0\equiv\varepsilon$. 

We can use (\ref{dnd}) to see that the OPE $\psi\cdot\sigma=\Phi_{1,3}\cdot\Phi_{1/2,0}\bar{\Phi}_{1/2,0}$ produces a field, let us call it $\mu$, with the same conformal dimensions of $\sigma$. This in turn leads to $\sigma\cdot\mu=\psi+\cdots$, meaning that $\sigma$ and $\mu$ cannot coincide\footnote{This is possible because the conditions for degenerate fields only determine the dimensions of the fields appearing in the OPE. Fields with the same dimensions may differ if this is required by consistence with mutual locality.} because they are mutually non-local. The existence of the field $\mu$ is not surprising once we recognize $\varepsilon_k\to(-1)^{k+1}\varepsilon_k$ as a duality transformation under which $\sigma\to\mu$; such a duality is well known for the particular case of the Ising model. We see that for $\varepsilon=\Phi_{2,1}\bar{\Phi}_{2,1}$ the notion of parafermion we introduced matches the usual one \cite{FrK,FZ} of field produced in the OPE of the spin field with its dual.

We conclude this section with the case $\varepsilon=\Phi_{1,3}\bar{\Phi}_{1,3}$. An analysis completely analogous to that of the previous case selects $\psi=\Phi_{2,1}$. The OPE $\varepsilon\cdot\varepsilon$ and iterations generate the subleading thermal fields $\varepsilon_k=\Phi_{1,2k+3}\bar{\Phi}_{1,2k+3}$, $k=1,2,\ldots$, without that a dual symmetry emerges in this case. As for the spin field, this time the condition (\ref{epsilonsigma}) is automatically satisfied and gives no constraint.

\section{Scattering and symmetries}
In the previous section we identified two critical lines of conformal field theory with central charge smaller than 1. The analysis was based on degenerate fields and their OPE's  and does not seem to allow the identification of internal symmetries in any obvious way. On the other hand, internal symmetries play an essential role within the scattering framework described in section~2. We now apply the scattering formalism to the cases of $O(n)$ and $S_q$ symmetries and match the results with those of the previous section.

\subsection{$O(n)$ symmetry and self-avoiding walks}
\begin{figure}
\begin{center}
\includegraphics[width=9cm]{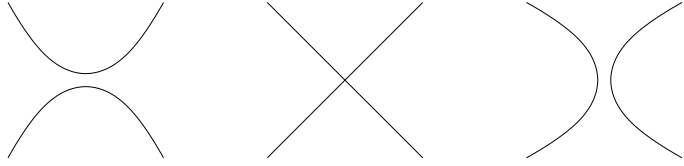}
\caption{Amplitudes $S_1$, $S_2$ and $S_3$ of the $O(n)$-invariant theory. Time runs vertically.}
\label{On_ampl}
\end{center} 
\end{figure}

It is natural to start with a multiplet of neutral excitations labeled by an index $a=1,\ldots,n$. The relation (\ref{algebra}) takes the $O(n)$-covariant form (Fig.~\ref{On_ampl})
\EQ
\Psi_a\circ\bar{\Psi}_b=\delta_{ab}\,S_1\sum_{c=1}^n\bar\Psi_c\circ\Psi_c+
S_2\,\,\bar\Psi_b\circ\Psi_a+S_3\,\,\bar\Psi_a\circ\Psi_b\,.
\label{On_algebra}
\EN
The unitarity equations read
\bea
& S_2S_2^*+S_3S_3^*=1\,,\\
& S_2S_3^*+S_3S_2^*=0\,,\\
& nS_1S_1^*+S_2S_1^*+S_1S_2^*+S_3S_1^*+S_1S_3^*=0\,,
\eea
while crossing symmetry gives
\EQ
S_1=S_3^*\equiv\rho_1e^{i\varphi}\,,\hspace{1cm}S_2=S_2^*\equiv\rho_2\,,
\EN
with $\rho_1$ non-negative and $\rho_2$ real. The unitarity equations are then rewritten as
\bea
&& \rho_1^2+\rho_2^2=1\,,
\label{uni1}\\
&& \rho_1\rho_2\cos\varphi=0\,,
\label{uni2}\\
&& n\rho_1^2+2\rho_1\rho_2\cos\varphi+2\rho_1^2\cos 2\varphi=0\,.
\label{uni3}
\eea
There are three ways of satisfying (\ref{uni2}). The first possibility is that $\cos\varphi=0$, so that (\ref{uni3}) fixes $n=2$; we will come back to this case in section~4.3. The second possibility is that $\rho_1=0$, but then we are left with $S_2=\pm 1$ as the only non-vanishing amplitude, i.e. with $n$ decoupled free bosons or fermions, an uninteresting case. Hence, the only non-trivial case with a continuous $n$-dependence is
\EQ
\rho_2=0\,,\hspace{1cm}\rho_1=1\,,\hspace{1cm}n=-2\cos 2\varphi\,.
\label{On}
\EN
It is natural to think of the particle trajectories as those of walks of $n$ different colors on the Euclidean plane. The result $S_2=0$ then amounts to a condition of self and mutual avoidance of the walks (Fig.~\ref{On_ampl}). This was first observed in \cite{selfavoiding} in the study of the off-critical case.

It follows from (\ref{On_algebra}) that 
\EQ
\sum_{a=1}^n\Psi_a\circ\bar{\Psi}_a=S\,\,\sum_{a=1}^n\bar\Psi_a\circ\Psi_a\,,
\label{algebra_On}
\EN
\EQ
S=nS_1+S_2+S_3=-e^{3i\varphi}\,,
\label{S_On}
\EN
where in the last equality we used the solution (\ref{On}). Equation (\ref{phase}) then relates $\varphi$ to $\Delta_\psi$, up to the $2\pi k$ ambiguity involved in the comparison of phases.

The solution (\ref{On}) covers the range $n\in(-2,2)$. The relations (\ref{On}) and (\ref{S_On}) consistently give $n=1$ and the fermionic (Ising) value $S=-1$ when $\varphi=2\pi k/3$, $k=1,2\,$(mod\,3). The $O(n)$-invariant critical line we are discussing should then be matched with one of the two critical theories with $c<1$ that we identified in the previous section. Since no order-disorder duality is expected for $O(n)$ symmetry, the natural choice is the theory with $\varepsilon=\Phi_{1,3}\bar{\Phi}_{1,3}$ and $\psi=\Phi_{2,1}$. We know that $\Delta_\varepsilon=\Delta_{1,3}=1$ at $c=1$ and that for $\Delta_\varepsilon=1$ (i.e. for $b^2=1$) on the $O(2)$ line the dimension of the parafermion is $\Delta_\psi=1/4$. With this further piece of information we fix $\varphi=-\frac{2\pi}{3}(\Delta_\psi+1/2)$\,(mod\,$2\pi$); substituting $\Delta_\psi=\Delta_{2,1}$ and plugging into (\ref{On}) we finally obtain
\EQ
n=2\cos\frac{\pi}{t}\,,
\label{n}
\EN
where $t$ was introduced in (\ref{beta-t}) and parameterizes the central charge.

We saw that the spin field has $\Delta_\sigma=1/16b^2$ along the $O(2)$ line with $c=1$, and that the $n=2$ endpoint of the solution (\ref{On}) corresponds to $b^2=1$. Hence we conclude that for the latter theory $\Delta_\sigma=1/16$ not only at $c=1/2$ but also at $c=1$; using these conditions to determine the indices (assumed $t$-independent) in (\ref{deltamunu}) gives $\Delta_\sigma=\Delta_{1/2,0}$ as the only positive solution within the range $c\in(0,1)$. The iteration of the OPE with $\varepsilon$ then produces subleading spin fields $\sigma_k$ with dimension $\Delta_{1/2,\pm 2k}$, $k=1,2,\ldots$.

\subsection{$S_q$ symmetry and cluster boundaries}
\begin{figure}
\begin{center}
\includegraphics[width=10cm]{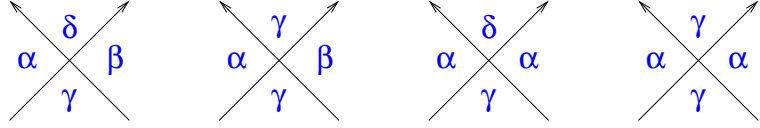}
\caption{Amplitudes $S_0$, $S_1$, $S_2$ and $S_3$ of the $S_q$-invariant theory.}
\label{potts_ampl}
\end{center} 
\end{figure}

The natural particle basis for an $S_q$-invariant theory below the critical temperature is that of the massive kinks corresponding to domain walls between the $q$ different ordered phases produced by spontaneous symmetry breaking \cite{CZ}. Such an interpretation of the excitations does not apply as it is to the critical case we want to investigate because at criticality the symmetry is unbroken and there is a single disordered phase. On the other hand, we can keep the same representation of the symmetry if we interpret the trajectories of the massless particles as the boundaries separating clusters of spins of different colors. We then denote by $\Psi_{\alpha\beta}$ ($\alpha,\beta=1,\ldots,q$; $\alpha\neq\beta$) the parafermion corresponding to the particle whose trajectory separates a cluster of color $\alpha$ from one of color $\beta$. Permutational symmetry of the colors now gives to (\ref{algebra}) the form (Fig.~\ref{potts_ampl})
\bea
\Psi_{\alpha\gamma}\circ\bar{\Psi}_{\gamma\beta}=(1-\delta_{\alpha\beta})
\left[S_0\sum_{\delta\neq\gamma}\bar\Psi_{\alpha\delta}\circ\Psi_{\delta\beta}+
S_1\,\,\bar\Psi_{\alpha\gamma}\circ\Psi_{\gamma\beta}\right]&&\nonumber\\
+ \delta_{\alpha\beta}\left[
S_2\sum_{\delta\neq\gamma}\bar\Psi_{\alpha\delta}\circ\Psi_{\delta\alpha}+  
S_3\,\,\bar\Psi_{\alpha\gamma}\circ\Psi_{\gamma\alpha}\right]&&\,,
\label{algebra_Sq}
\eea
where it is understood that the two indices of a parafermionic field have to be different.

Crossing symmetry leads to the relations
\EQ
S_0=S_0^*\equiv\rho_0\,,\hspace{1cm}S_1=S_2^*\equiv\rho e^{i\varphi}\,,\hspace{1cm}S_3=S_3^*\equiv\rho_3\,,
\EN
with $\rho_0$ and $\rho_3$ real and $\rho$ non-negative. With these parameterizations the unitarity equations read
\bea
&&(q-3)\rho_0^2+\rho^2=1\,,\label{unit1}\\
&&(q-4)\rho_0^2+2\rho_0\rho\cos\varphi=0\,,\label{unit2}\\
&&(q-2)\rho^2+\rho_3^2=1\,,\label{unit3}\\
&&(q-3)\rho^2+2\rho\rho_3\cos\varphi=0\,.\label{unit4}
\eea
It is straightforward to see that they produce a quadratic equation for $\rho^2$ with solutions $\rho^2_{(1)}=4-q$ and $\rho^2_{(2)}=(q-4)/(q^2-5q+5)$. Choosing the first one leads to
\EQ
\rho_0=-1\,,\hspace{1cm}\rho=\sqrt{4-q}\,,\hspace{1cm}2\cos\varphi=-\sqrt{4-q}\,,\hspace{1cm}\rho_3=q-3\,,
\label{potts}
\EN
provided we use the condition $\rho_3=-1$ at $q=2$ (Ising) to fix the overall sign of $\rho_3$. It follows from the form of the solution (\ref{potts}) that it holds for $q\in(0,4)$.

\begin{figure}
\begin{center}
\includegraphics[width=9cm]{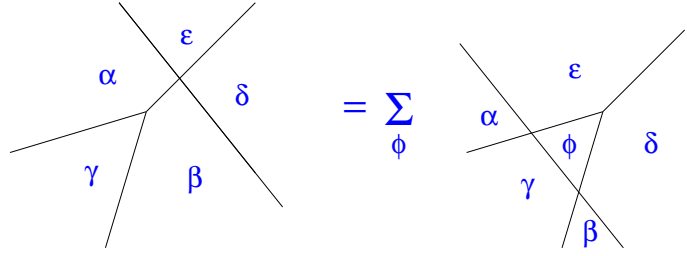}
\caption{Diagrammatic equation which for different choices of external indices produces (\ref{boot_i}-\ref{boot_f}).}
\label{paraf_bootstrap}
\end{center} 
\end{figure}

An interesting way to see that (\ref{potts}) is the solution we are interested in is to observe that a cluster boundary can branch into two boundaries at a point where three clusters meet, so that we expect the fusion relation
\EQ
\Psi_{\alpha\gamma}\cdot\Psi_{\gamma\beta}\sim\Psi_{\alpha\beta}\,,
\EN
with a structure constant $C_{\Psi\Psi}^\Psi$ which is color-independent by permutational symmetry. The requirement that the result of the scattering of $\bar{\Psi}_{\beta\delta}$ with the r.h.s. equals that of the scattering with the l.h.s. yields the equations (Fig.~\ref{paraf_bootstrap})
\bea
S_1=S_2^2+(q-3)S_2S_0\,,&\hspace{1cm}&S_0=S_3S_0+S_2^2+(q-4)S_2S_0\,,
\label{boot_i}\\
S_3=(q-3)S_0^2\,,&  &S_2=S_1^2+(q-3)S_0S_1\,,\\
S_2=S_1S_0+(q-4)S_0^2\,,&  &S_1=S_0S_2+(q-4)S_0^2\,,\\
S_0=S_1^2+S_0S_3+(q-4)S_0S_1\,,&  &S_0=S_1S_0+S_0S_2+(q-5)S_0^2\,,
\label{boot_f}
\eea
whose unique non-vanishing solution is (\ref{potts}).

At $q=2$ the only physical amplitude is $S_3=-1$, as required for the Ising model. The solution (\ref{potts}) then corresponds to a critical line spanning a neighborhood of $c=1/2$, and it is naturally associated to the remaining theory with $c\leq 1$ of the previous section, namely the one with $\varepsilon=\Phi_{2,1}\bar{\Phi}_{2,1}$ and $\psi=\Phi_{1,3}$; the order-disorder duality that we found in such a theory is known to hold for the two-dimensional Potts model. The maximal value of central charge, $c=1$, should correspond to the maximal value of $q$, which we saw is 4.

It follows from (\ref{algebra_Sq}) that
\EQ
\sum_\gamma\Psi_{\alpha\gamma}\circ\bar{\Psi}_{\gamma\alpha}=
S\,\,\sum_\gamma\bar\Psi_{\alpha\gamma}\circ\Psi_{\gamma\alpha}\,,
\EN
\EQ
S=S_3+(q-2)S_2=e^{-4i\varphi}\,,
\label{S_Sq}
\EN
with the last equality produced by the solution (\ref{potts}). Comparison with (\ref{phase}), together with the conditions that $\Delta_\psi=\Delta_{1,3}$ equals $1/2$ and 1 for $q=2$ and 4, respectively, determine $\varphi=\pi(1+\frac12\Delta_{1,3})$ (mod\,$2\pi$), which substituted in (\ref{potts}) leads to the relation between $q$ and the central charge
\EQ
\sqrt{q}=2\sin\frac{\pi(t-1)}{2(t+1)}\,.
\label{q}
\EN

\subsection{$O(2)$ symmetry and sine-Gordon coupling}
We saw that (\ref{uni1}-\ref{uni3}) admit a solution with $n=2$. It can be parameterized as
\EQ
\rho_1=\sin\alpha\,,\hspace{1cm}\rho_2=\cos\alpha\,,\hspace{1cm}\varphi=-\pi/2\,,\label{O2}
\EN
where we now allow for $\rho_1$ negative. The phase in (\ref{algebra_On}) becomes
\EQ
S=2S_1+S_2+S_3=e^{-i\alpha}\,,
\label{S_O2}
\EN
and has to be equated to (\ref{phase}) with $\Delta_\psi$ which, as seen in the previous section, takes the value $1/4b^2$ along the $O(2)$ critical line; this gives 
\EQ
\alpha=\frac{\pi}{2b^2}\,.
\label{alpha}
\EN

The $S$-matrix of a massive $O(2)$-invariant integrable theory was determined in \cite{ZZ} and found to depend on a non-negative parameter $\xi$. This $S$-matrix corresponds to the sine-Gordon model obtained perturbing the $O(2)$-invariant critical line with $c=1$ by the field $\varepsilon=\cos 2b\varphi$, and $\xi$ is a function of $b$; the $S$-matrix becomes minus the identity when $\xi=\pi$, which then corresponds to the free fermion point $b^2=1/2$. The elementary excitations are solitons and anti-solitons interpolated by the fields $\Psi_\pm=\Psi_1\pm i\Psi_2$. As a consequence, the asymptotic high energy limit of the soliton-antisoliton amplitude, which turns out to be $e^{-i\frac{\pi}{2}(1+\frac{\pi}{\xi})}$ \cite{ZZ}, should coincide with the massless amplitude (\ref{S_O2}). Using (\ref{alpha}) and the condition at $\xi=\pi$ one obtains the relation
\EQ
\xi=\frac{\pi b^2}{1-b^2}\,,
\label{xi}
\EN
which was introduced in \cite{ZZ} through semiclassical and perturbative arguments.

\begin{table}
\begin{center}
\begin{tabular}{|l|c|c|c|c|}
\hline
Symmetry & $c$ & $\Delta_{\varepsilon_k}$ & $\Delta_\psi$ & $\Delta_{\sigma_k}$ \\
\hline
$O(2)$ & $1$ & $b^2$ & $\frac{1}{4b^2}$ & $\frac{1}{16b^2}$ \\
$O(n)\hspace{.4cm}n=2\cos\frac{\pi}{t}$ & $1-\frac{6}{t(t+1)}$ & $\Delta_{1,2k+3}$ & $\Delta_{2,1}$ & $\Delta_{1/2,\pm 2k}$ \\
$S_q\hspace{.4cm}\sqrt{q}=2\sin\frac{\pi(t-1)}{2(t+1)}$ & $1-\frac{6}{t(t+1)}$ & $\Delta_{k+2,1}$ & $\Delta_{1,3}$ & $\Delta_{k+1/2,0}$ \\
\hline
\end{tabular}
\caption{Central charge and conformal dimensions along the three critical lines analyzed in the paper. $\Delta_\psi$ is the parafermionic dimension, while $\Delta_{\varepsilon_k}$ and $\Delta_{\sigma_k}$ are the thermal and magnetic dimensions, respectively, with the leading dimensions (the only ones quoted for $O(2)$) corresponding to $k=0$ and the subleading ones corresponding to $k=1,2,\ldots$; $\Delta_{\mu,\nu}$ is given in (\ref{deltamunu}).} 
\label{table1}
\end{center}
\end{table}

\section{Conclusion} 
The relations (\ref{n}) and (\ref{q}) determine the $n$- and $q$-dependence of the conformal dimensions identified in section~3 and at the end of section~4.1; the results are summarized in Table~1. The values of the leading and some subleading dimensions $\Delta_{\varepsilon_k}$ and $\Delta_{\sigma_k}$ obtained within the lattice Coulomb gas framework \cite{Nienhuis} can be checked to coincide with those in Table~1. The latter also gives the information about the $O(2)$-symmetric critical line. It was shown in section~4.3 how the formalism of this paper allows for a simple derivation of the relation (\ref{xi}) between the coordinate $b$ along this line and the parameter $\xi$ entering the $S$-matrix of the massive $O(2)$-invariant theory (sine-Gordon model).

Other symmetries can be considered. For example, ${\bf Z}_N$ is implemented on a basis of parafermions $\Psi_a$, with $a=1,2,\ldots,N-1$ denoting the ${\bf Z}_N$ charge and $\Delta_{\Psi_a}=\Delta_{\Psi_{N-a}}$ by charge conjugation. The diagonal solution $S_{ab}^{cd}=\delta_a^c\delta_b^dS_{ab}$ is subject to the unitarity and crossing constraints $S_{ab}=1/S^*_{ab}=S^*_{a,N-b}$. Requiring the fusion $\Psi_a\cdot\Psi_b\sim\Psi_{a+b\,(mod\,N)}$ leads to $S_{ac}S_{bc}=S_{a+b,c}$ and to the result $S_{ab}=e^{2i\pi ab/N}$, which in turn gives $(i/2\pi)\ln S_{aa}=-a^2/N=\Delta_{\Psi_a}$\,(mod\,1). The requirements that $\Delta_{\Psi_1}=1/2$ in the Ising case $N=2$ and that for $N=3$ $\Delta_{\Psi_1}=\Delta_{\Psi_2}$ takes the value $2/3$ that we obtained above for $S_3$ symmetry lead to $\Delta_{\Psi_a}=a(N-a)/N$. This value is a starting point of the construction of ${\bf Z}_N$-invariant conformal field theories in \cite{FZ}.

This work originates from a study concerning the space of fields of massive field theories \cite{D09} where, in particular, it was observed that parafermionic fields naturally appear in the analysis of high energy asymptotics (see also \cite{Smirnov}). In the present paper, working directly at criticality, we have seen how particle excitations are described by chiral fields which in general have parafermionic nature, and how non-trivial results, in particular for non-rational conformal field theories, can be obtained studying parafermions in both the operator product and scattering settings. It is interesting to notice that in recent years discrete parafermionic observables were found to provide an intriguing link between criticality, holomorphicity and integrability on the lattice (see \cite{Cardy09} for a review). The universal role of parafermions exhibited in this paper directly in the continuum may also be relevant for the interpretation of some of the lattice findings.




\begin{thebibliography}{99}
\bibitem{BPZ}  A.A. Belavin, A.M. Polyakov and A.B. Zamolodchikov, Nucl. Phys. B 241 (1984) 333.

\bibitem{Baxter} R.J. Baxter, Exactly Solved Models of Statistical Mechanics, Academic Press, London, 1982.

\bibitem{FK}  C.M. Fortuin and P.W. Kasteleyn, J. Phys. Soc. Jpn. Suppl. 26 (1969) 11; Physica 57 (1972) 536.

\bibitem{deGennes} P.G. de Gennes, Phys. Lett. A 38 (1972) 339.

\bibitem{Nienhuis} B. Nienhuis, Coulomb gas formulation of two-dimensional phase transitions, in C. Domb and J.L. Lebowitz (eds.), Phase transitions and critical phenomena, vol. 11, p. 1-53, Academic Press, London, 1987.

\bibitem{Dotsenko} Vl.S. Dotsenko, Nucl. Phys. B 235 (1984) 54.

\bibitem{DF} Vl.S. Dotsenko and V.A. Fateev, Nucl. Phys. B 240 (1984) 312.

\bibitem{Coleman} S. Coleman, Phys. Rev. D 11 (1975) 2088.

\bibitem{Mandelstam} S. Mandelstam, Phys. Rev. D 11 (1975) 3026.

\bibitem{FrK} E. Fradkin and L.P. Kadanoff, Nucl. Phys. B 170 (1980) 1.

\bibitem{FZ} A.B. Zamolodchikov and V.A. Fateev, Sov. Phys. JETP 62 (1985) 215.

\bibitem{selfavoiding} A.B. Zamolodchikov, Mod. Phys. Lett. A 6 (1991) 1807.

\bibitem{CZ} L. Chim and A.B. Zamolodchikov, Int. J. Mod. Phys. A 7 (1992) 5317.

\bibitem{ZZ} A.B. Zamolodchikov and Al.B. Zamolodchikov, Annals of Physics 120 (1979) 253.

\bibitem{D09} G. Delfino, Nucl. Phys. B 807 (2009) 455.

\bibitem{Smirnov} F. Smirnov, Commun. Math. Phys. 132 (1990) 415.

\bibitem{Cardy09} J. Cardy, J. Stat. Phys. 137 (2009) 814.



\end{thebibliography}
\end{document}